\documentclass[a4paper,12pt]{article}
\usepackage{amsfonts,amstext,latexsym,times,xspace}
\usepackage{epsfig}

\newcommand{\lb}[1]{\label{#1}}

\renewcommand{\[}{\begin{eqnarray}}
\renewcommand{\]}{\end{eqnarray}}
\newcommand{\nn}{\nonumber}
\newcommand{\non}{\nonumber \\ }
\renewcommand{\=}{\equiv}
\def\ba{\begin{array}}
\def\ea{\end{array}}
\makeatletter
\newif\if@fewtab\@fewtabtrue
\def\moth{\mathsurround=0pt}
\newdimen\zo \zo=0pt

\def\tick{\leaders\hrule height 0.5ex depth 0pt \hskip 0.5pt}
\def\upboxfill{$\moth \setbox\zo\hbox{\tick}%
  \hskip 2pt\hbox to 0pt{$\tick$\hss}\hrulefill \hbox to 2pt{$\tick$\hss}$}
\def\underbox#1{\offinterlineskip{\mathord{\mathop{\vtop{\moth\ialign{##\crcr
      $\hfil\displaystyle{#1}\hfil$\crcr\noalign{}
      {\upboxfill}\crcr\noalign{}}}}\limits}}}
\def\dtick{\leaders\hrule height .34pt depth 0.5ex \hskip 0.5pt}
\def\downboxfill{$\moth \setbox\zo\hbox{\dtick}%
  \hskip 2pt\hbox to 0pt{$\dtick$\hss}\hrulefill%
  \hbox to 2pt{$\dtick$\hss}$}
\def\overbox#1{\mathop{\vbox{\moth\ialign{##\crcr\noalign{}
\downboxfill\crcr\noalign{\vskip 1pt\nointerlineskip}
      $\hfil\displaystyle{#1}\hfil$\crcr}}}\limits}

\newcommand{\undersym}[1]{\underbox{{}#1}}
\newcommand{\oversym}[1]{\!\overbox{{}#1}}

\newcommand{\cD}{\ensuremath{\mathcal{D}}\xspace}

\newcommand{\6}[1][{}]{\ensuremath{E_{6 #1}}\xspace}
\newcommand{\7}[1][{}]{\ensuremath{E_{7 #1}}\xspace}

\newcommand{\E}{E_{8(8)}}
\newcommand{\EE}{E_{7(7)}}
\newcommand{\EEE}{E_{6(6)}}

\newcommand{\USp}[1][8]{\ensuremath{\mbox{USp$(#1)$}}\xspace}

\newcommand{\SLR}[1][8]{\ensuremath{\mbox{SL$(#1,\mathbb{R})$}}\xspace}

\newcommand{\fg}{\ensuremath{\mathfrak{g}}\xspace}

\newcommand{\gO}[1][{}]{\Omega^{#1}\xspace}
\newcommand{\gOd}[1][{}]{\Omega_{#1}\xspace}

\newcommand{\gd}[1][{}]{\delta_{#1}{}}

\newcommand{\eps}[1][{}]{\epsilon^{#1}\xspace}
\newcommand{\epsd}[1][{}]{\epsilon_{#1}\xspace}

\newcommand{\Zt}{\tilde{Z}{}}
\newcommand{\Zbt}{\bar{Z}{}}

\newcommand{\Et}{\tilde{E}}
\newcommand{\Ht}{\tilde{H}}
\newcommand{\Ft}{\tilde{F}}
\newcommand{\Gt}{\tilde{G}}
\newcommand{\Xt}{\tilde{X}}
\newcommand{\Yt}{\tilde{Y}}

\newcommand{\ft}[2]{{\textstyle {\frac{#1}{#2}} }}

\newcommand{\st}{^\ast}

\newcommand{\rep}[1]{\ensuremath{\mbox{\mathversion{bold}$\mathbf{#1}$%
                     \mathversion{normal}}}}

\renewcommand{\i}{\mathrm{i}}

\newcommand{\cX}{\mathcal{X}}
\newcommand{\cY}{\mathcal{Y}}

\newcommand{\cN}{\mathcal{N}}
\newcommand{\cI}{\mathcal{I}}

\newcommand{\Com}[2]{[#1\, ,\,#2]}
\newcommand{\Sympl}[2]{\left\langle #1,#2\right\rangle}

\newcommand{\Rn}{\ensuremath{\mathbb{R}}\xspace}

\begin{document}
\title{Realizations of Exceptional  U-duality Groups as Conformal and  Quasi-conformal Groups and  Their Minimal
Unitary  Representations}

\author{Murat Gunaydin\thanks{ Work supported in part by the
    National Science Foundation under grant number PHY-0245337.}\\ Physics Department , Penn State University \\ University Park, PA 16802 \\ e-mail:
murat@phys.psu.edu}
\date{~}
 \maketitle

\begin{abstract}
{\small
 We review the novel quasiconformal realizations of exceptional U-duality groups whose "quantization" lead
directly to their minimal unitary irreducible representations. The group $E_{8(8)}$ can be realized as a
quasiconformal group in the 57 dimensional charge-entropy space of BPS black hole solutions of maximal $N=8$
supergravity in four dimensions and leaves invariant "lightlike separations" with respect to a quartic norm.
Similarly $E_{7(7)}$ acts as a conformal group in the 27 dimensional charge space of BPS black hole solutions in
five dimensional $N=8$ supergravity and leaves invariant  "lightlike separations" with respect to a cubic norm.
For the exceptional $N=2$ Maxwell-Einstein supergravity theory the corresponding quasiconformal and conformal
groups are $E_{8(-24)}$ and $E_{7(-25)}$, respectively. These conformal and quasiconformal groups act as spectrum
generating symmetry groups in five and four dimensions and are isomorphic to the U-duality groups of the
corresponding  supergravity theories in four and three dimensions, respectively. Hence the spectra of these
theories are expected to form unitary representations of these groups whose minimal unitary realizations are also
reviewed.}
\end{abstract}
\section{U-Duality Groups in Supergravity Theories}
\subsection{Noncompact exceptional groups as symmetries of maximally extended supergravity theories}
Eleven dimensional supergravity \cite{crjusc} is the effective low energy theory of strongly coupled phase of
M-theory \cite{witten95}. Toroidal compactification of eleven dimensional supergravity theory down to $d$
dimensions yields the maximally extended supergravity with a global non-compact symmetry group $E_{(11-d)(11-d)}$
\cite{julia}. It is believed that only the discrete subgroups $E_{(11-d)(11-d)}(\mathbb{Z})$ of these groups are
the symmetries of the non-perturbative spectra of toroidally compactified  M-theory \cite{huto}. We shall use the
term U-duality for these discrete subgroups as well as for the global noncompact symmetry groups of supergravity
theories.

In five dimensions  $E_{6(6)}$ is a symmetry of the Lagrangian of the maximal ( $N=8$) supergravity, under which
the 27 vector fields of the theory transform  irreducibly while the 42 scalar fields   transform nonlinearly and
parameterize the coset space $E_{6(6)} / USp(8)$. On the other hand the $E_{7(7)}$ symmetry of the maximally
extended supergravity in $d=4$ is  an on-shell symmetry group. The field strengths of  28 vector fields of this
theory together with their "magnetic"  duals transform irreducibly in the 56 of $E_{7(7)}$ and  70 scalar fields
parameterize the coset space $E_{7(7)}/SU(8)$. In three dimensions all the propagating bosonic degrees of the
maximal $N=16$ supergravity can be dualized to scalar fields, which transform nonlinearly under the  symmetry
group $E_{8(8)}$ and parameterize the coset space $E_{8(8)}/SO(16)$.

\subsection{Exceptional U-duality groups in Matter coupled Supergravity Theories}
Non-compact global U-duality groups arise in matter coupled supergravity theories as well. In this talk I will
focus on U-duality groups in $N=2$ Maxwell-Einstein supergravity theories (MESGT) in $d=5$ and the corresponding
theories in four and three dimensions. These theories describe the coupling of an arbitrary number $n$ of (
Abelian) vector fields to $N=2$ supergravity and five dimensional  MESGT's were constructed in \cite{GST1}. The
bosonic part of the Lagrangian is given by \cite{GST1}
\begin{eqnarray}\label{Lagrange}
e^{-1}\mathcal{L}_{\rm bosonic}&=& -\frac{1}{2}R -\frac{1}{4}{\stackrel{\circ}{a}}_{IJ}F_{\mu\nu}^{I}
F^{J\mu\nu}-\frac{1}{2}g_{xy}(\partial_{\mu}\varphi^{x}) (\partial^{\mu} \varphi^{y}) \nonumber \\ &&+
 \frac{e^{-1}}{6\sqrt{6}}C_{IJK}\varepsilon^{\mu\nu\rho\sigma\lambda}
 F_{\mu\nu}^{I}F_{\rho\sigma}^{J}A_{\lambda}^{K},
\end{eqnarray}
where  $e$ and $R$ denote the f\"{u}nfbein determinant and the scalar curvature, respectively, and
$F_{\mu\nu}^{I}$ are the field strengths of the Abelian vector fields $A_{\mu}^{I} , \,( I=0,1,2 \cdots , n$).
The metric, $g_{xy}$, of the scalar manifold $\mathcal{M}$  and the "metric"  ${\stackrel{\circ}{a}}_{IJ}$ of the
kinetic energy term of the vector fields both depend on the scalar fields $\varphi^{x}$. The Abelian gauge
invariance requires the completely symmetric tensor $C_{IJK}$ to be  constant. Remarkably, the entire $N=2$,
$d=5$ MESGT  is uniquely determined by the constant tensor $C_{IJK}$ \cite{GST1}. To see this explicitly for the
bosonic terms in the Lagrangian consider  the  cubic polynomial, $\mathcal{V}(h)$,
 in $(n+1)$
real variables $h^{I}$ $(I=0,1,\ldots,n)$ defined by the $C_{IJK}$
\begin{equation}
\mathcal{V}(h):=C_{IJK}h^{I}h^{J}h^{K}\ .
\end{equation}
Using this polynomial as a real "potential" for a metric, $a_{IJ}$,
 in the (ambient) space $\mathbb{R}^{(n+1)}$ with the coordinates  $h^{I}$:
\begin{equation}\label{aij}
a_{IJ}(h):=-\frac{1}{3}\frac{\partial}{\partial h^{I}} \frac{\partial}{\partial h^{J}} \ln \mathcal{V}(h)\ .
\end{equation}
one finds that the   $n$-dimensional    target space, $\mathcal{M}$, of the scalar fields $\varphi^{x}$ can then
be represented as the hypersurface \cite{GST1}
\begin{equation}
\mathcal{V} (h)=C_{IJK}h^{I}h^{J}h^{K}=1
\end{equation}
in this ambient space. The metric  $g_{xy}$ is simply  the pull-back of (\ref{aij}) to $\mathcal{M}$ and the
"metric" ${\stackrel{\circ}{a}}_{IJ}(\varphi)$ appearing in (\ref{Lagrange}) is given by the componentwise
restriction of $a_{IJ}$ to $\mathcal{M}$:
\[
{\stackrel{\circ}{a}}_{IJ}(\varphi)=a_{IJ}|_{\mathcal{V}=1} \ .
\]

The positivity of kinetic energy  requires that  $g_{xy}$ and ${\stackrel{\circ}{a}}_{IJ}$   be positive
definite. This requirement induces constraints on the possible $C_{IJK}$, and in \cite{GST1} it   was shown that
any $C_{IJK}$ that satisfy these constraints can be brought to the following form
\begin{equation}\label{canbasis}
C_{000}=1,\quad C_{0ij}=-\frac{1}{2}\delta_{ij},\quad  C_{00i}=0,
\end{equation}
with  the remaining coefficients $C_{ijk}$
 ($i,j,k=1,2,\ldots , n$) being  completely arbitrary.
 This basis is referred to as the canonical basis for
$C_{IJK}$.

Denoting the symmetry group of the tensor $C_{IJK}$ as $G$ one finds that  the full symmetry group of  $N=2$
MESGT in $d=5$ is of the form $G \times SU(2)_R$ , where $SU(2)_R$ denotes the local R-symmetry group of the
$N=2$ supersymmetry algebra. A MESGT is said to be {\it unified} if all the vector fields, including the
graviphoton, transform in an irreducible representation of a {\it simple}  symmetry group $G$. Of all the $N=2$
MESGT's whose scalar manifolds are symmetric spaces only four are unified \cite{GST1,GST3}\footnote{ It has
recently been shown that if one relaxes the condition that the scalar manifolds be homogeneous spaces then  there
exist three novel infinite families ( plus an additional sporadic one ) of   unified MESGT's in $d=5$
\cite{gz2003}. }. If one defines a cubic form $ \mathcal{N}(h):=C_{IJK}h^{I}h^{J}h^{K} $ using the constant
tensor $C_{IJK}$ , it was shown in \cite{GST1} that the cubic forms associated with the four unified MESGT's can
be identified with the norm forms of simple ( Euclidean) Jordan algebras of degree three . There exist only four
simple ( Euclidean) Jordan algebras of degree three and they can be realized in terms of $3\times 3$ hermitian
matrices over the four division algebras with the product being one-half the anticommutator. They are denoted as
$J_3^{\mathbb{A}}$, where $\mathbb{A}$ stands for the underlying division algebra , which can be real numbers
$\mathbb{R}$, complex numbers $\mathbb{C}$, quaternions $\mathbb{H}$ and octonions $\mathbb{O}$. The supergravity
theories defined by them were referred to as magical supergravity theories \cite{GST1} since their symmetry
groups in five , four and three dimensions correspond precisely to the symmetry groups of the famous Magic Square
. The octonionic Jordan algebra $J_3^{\mathbb{O}}$ is  the unique exceptional Jordan algebra and consequently the
$N=2$ MESGT defined by it is called the exceptional supergravity theory \cite{GST1}. In the table below we list
the scalar manifolds $G/H$ of the magical supergravity theories in five , four and three dimensions, where $G$ is
the global noncompact symmetry and $H$ is its maximal compact subgroup.

\begin{center}
\begin{small}
\begin{tabular}{|c|c|c|c|}
\hline $J$ & G/H in d=5 & G/H in d=4 & G/H in d=3 \\ \hline
 $J_3^{\mathbb{R}}$ & $SL(3,\mathbb{R})/SO(3)$ & $Sp(6,\mathbb{R})/U(3)$ &$F_{4(4)}/USp(6)\times SU(2)$ \\
 \hline
$ J_3^{\mathbb{C}} $ &$SL(3,\mathbb{C})/U(3)$&$SU(3,3)/SU(3)^2 \times U(1) $ &$E_{6(2)}/SU(6)\times SU(2) $ \\
\hline $J_3^{\mathbb{H}}$ & $SU^*(6)/USp(6)$ &$SO^*(12) /U(6)$ &$E_{7(-5)}/SO(12)\times SU(2)$  \\ \hline
$J_3^{\mathbb{O}} $ & $E_{6(-26)}/F_4$ & $E_{7(-25)}/E_6 \times U(1)$ & $E_{8(-24)}/E_7 \times SU(2) $  \\ \hline

\end{tabular}

\end{small}
\end{center}

 Note that the exceptional $N=2$ supergravity has  $ E_{6(-26)}, E_{7(-25)}$ and $E_{8(-24)}$ as
  its global symmetry group in five , four and three dimensions, respectively,
 whereas the maximally extended supergravity theory has the maximally split real forms $E_{6(6)}, E_{7(7)}$ and $E_{8(8)}$
 as its symmetry in the respective dimensions.

The term U-duality was introduced by Hull and Townsend since the discrete symmetry group $E_{7(7)}(\mathbb{Z})$
of M/superstring theory toroidally compactified to $d=4$ {\it unifies} the T-duality group $SO(6,6)(\mathbb{Z})$
with the S-duality group $SL(2,\mathbb{Z})$ in a simple group:
\[ SO(6,6) \times SL(2,\mathbb{R}) \subset E_{7(7)} \]
The analogous decomposition of the symmetry group $E_{7(-25)}$ of the exceptional supergravity in $d=4$ is
\[ SO(10,2) \times SL(2,\mathbb{R}) \subset E_{7(-25)} \]
with similar decompositions for the other magical supergravity theories.

\section{U-duality Groups and black Hole Entropy in Supergravity Theories}

The entropy of  black hole solutions in maximally extended supergravity as well as in matter coupled supergravity
theories are invariant under the corresponding U-duality groups. For example in $d=5$ , $N=8$ supergravity the
entropy $S$  of a black hole solution can be written in the form \cite{d5bh}

\begin{equation}
S= \alpha \sqrt{I_3} =\alpha  \sqrt{ C_{IJK} q^Iq^Jq^K}
\end{equation}
where $\alpha$ is some fixed constant and $I_3$ is the cubic invariant of $E_{6(6)}$. The $q^I , I=1,2,...,27$
are the charges coupling to the 27 vector fields of the theory. The BPS black hole solutions with $I_3\neq 0$
preserve 1/8 supersymmetry \cite{fema}. The solutions with $I_3=0$ , but with $\frac{1}{3} \partial_I I_3 =
C_{IJK} q^Jq^K \neq 0$ preserve 1/4 supersymmetry , while those solution with both $I_3=0$ and  $C_{IJK}
q^Jq^K=0$ preserve 1/2 supersymmetry \cite{fema}. The orbits of the black hole solutions of $N=8$ supergravity in
$d=5$ under the action of $E_{6(6)}$ were classified in \cite{fegu}.

The entropy $S$ of black hole solutions of  $N=8$ supergravity in $d=4$ is given by the quartic invariant of the
U-duality group $E_{7(7)}$ \cite{d4bh}
\begin{equation}
S= \beta \sqrt{I_4} =\beta \sqrt{d_{IJKL} q^Iq^Jq^Kq^L}
\end{equation}
where $\beta$ is a fixed constant and $q^I , I=1,2,...,56$ represent the 28 electric and 28 magnetic charges. The
orbits of the BPS black hole solutions of $N=8$ supergravity preserving 1/8, 1/4 and 1/2 supersymmetry under the
action of $E_{7(7)}$ were classified in \cite{fegu}. The number of supersymmetries preserved by the extremal
black hole solutions depend on whether or not $I_4$ , $\partial_{J}I_4$ and $\partial_{J}
\partial_{K} I_4 $ vanish \cite{fema}.

The orbits of the BPS black hole solutions of $N=2$ MESGT's in $d=5$ and
 $d=4$ with symmetric target spaces , including those of the exceptional supergravity, were also classified in \cite{fegu}.

The classification of the orbits of BPS black hole solutions of the $N=8$ and the exceptional $N=2$ theory in
$d=5$ as given in \cite{fegu} associates with a given BPS black hole solution with charges $q^I$ an element
$J=\sum_{I=1}^{27} e_Iq^I $ of the corresponding exceptional Jordan algebra with basis elements $e_I$. ( Split
exceptional Jordan algebra $J_3^{\mathbb{O}_s}$ for the $N=8$ theory and the real exceptional Jordan algebra
$J_3^{\mathbb{O}}$ for the $N=2$ theory. )  The cubic invariant $I_3(q^I)$ is then simply given by the norm form
$\mathcal{N}$  of the Jordan algebra \[ \mathcal{N}_3(J) = I_3(q^I) \]
 Invariance group of the norm form ( known as the reduced structure group in mathematics
literature) is isomorphic to the U-duality group in these five dimensional theories. In the corresponding four
dimensional supergravity theory the black hole solution is associated with an element of the Freudenthal triple
system defined by the exceptional Jordan algebra and the U-duality group is isomorphic to the invariance group of
its associated quartic form \cite{fegu}.

\section{ Conformal and quasi-conformal realizations of exceptional groups}
The linear fractional group of the exceptional Jordan algebra $J_3^{\mathbb{O}} ( J_3^{\mathbb{O}_s} )$  as
defined by Koecher  is the exceptional group  $E_{7(-25)} ( E_{7(7)}) $ which can be interpreted as a generalized
conformal group of a "spacetime" coordinatized by $J_3^{\mathbb{O}} ( J_3^{\mathbb{O}_s} )$ \cite{seegkn}. Acting
on an element $J=\sum_{I=1}^{27} e_Iq^I $ the conformal action of $E_7$ changes its norm and hence the entropy of
the corresponding black hole solution. Thus one can regard $E_{7(-25)} ( E_{7(7)}) $ as a spectrum generating
symmetry in the charge space of black hole solutions of the exceptional $N=2$ ( $N=8$) supergravity in five
dimensions. If one defines a distance function between any two solutions with charges $q^I$ and $q'^{I}$ as \[
d(q,q') \equiv \mathcal{N}_3(J-J') \] one finds that the light like separations are preserved under the conformal
action of $E_7$ \cite{gkn1,gupa}. The explicit action of $E_{7(7)}$ and $E_{7(-25)}$  on the corresponding 27
dimensional spaces are given in \cite{gkn1} and \cite{gupa} , respectively. Let us review briefly the conformal
action of $\EE$ given in \cite{gkn1}.  Lie algebra of $\EE$ has a 3-graded decomposition
\[
\rep{133} = \rep{27}\oplus(\rep{78}\oplus\rep{1})\oplus\rep{\overline{27}}
\]
under its $\EEE \times \cD$ subgroup, where \cD represents the dilatation group $SO(1,1)$. Under its maximal
compact subalgebra \USp  Lie algebra \6[(6)] decomposes  as a symmetric tensor $\Gt^{ij}$ in the adjoint \rep{36}
of \USp and a fully antisymmetric symplectic traceless tensor $\Gt^{ijkl}$ transforming as the \rep{42} of \USp (
indices $1\le i,j,\ldots \le 8$ are  \USp indices). $\Gt^{ijkl}$ is traceless with respect to the real symplectic
metric $\gOd[ij]\!=\!-\gOd[ji]\!=\!-\gO[ij]$ (thus $\gOd[ik] \gO[kj]\!=\!\delta_i^j$). The symplectic metric is
used  to raise and lower indices, with the convention that this is always to be done from the left. The other
generators of conformal $\EE$ consist of  a dilatation generator $\Ht$, translation generators $\Et^{ij}$ and the
nonlinearly realized "special conformal" generators $\Ft^{ij}$, transforming as $\rep{27}$ and
$\rep{\overline{27}}$, respectively.

The fundamental \rep{27} of \6[(6)] on which  \7[(7)]) acts nonlinearly  can be represented as the  symplectic
traceless antisymmetric tensor $\Zt^{ij}$ transforming as \footnote{ Throughout we use the convention that
indices connected by a bracket are antisymmetrized with weight one.}
\[
\Gt^i{}_j(\Zt^{kl})     &=& 2\,\gd\oversym{ {}^k_j \Zt^{il}} \,, \non[1ex] \Gt^{ijkl}(\Zt^{mn})    &=&
\ft{1}{24}\eps[ijklmnpq]\Zt_{pq}\,,
 \lb{27-rep}
\]
where $ \Zt_{ij} \;:=\; \gOd[ik]\gOd[jl] \Zt^{kl} = (\Zt^{ij})\st\, \quad \text{ and } \quad \gOd[ij]\Zt^{ij}
\;=\; 0 $.
 The conjugate \rep{\overline{27}} representation transforms as
\[
\Gt^i{}_j(\Zbt^{kl})    &=& 2\,\gd\oversym{ {}^k_j \Zbt^{il}} \,, \non[1ex] \Gt^{ijkl}(\Zbt^{mn})   &=&
-\ft{1}{24}\eps[ijklmnpq]\Zbt_{pq}\,. \lb{27b-rep}
\]
 The cubic invariant of \6[(6)] in the \rep{27} is  given by
\[
  \cN_3 (\Zt) := \Zt^{ij} \Zt_{jk} \Zt^{kl} \gOd[il]\,.
\]

 The generators  $\Et^{ij}$ act as  translations on the space with coordinates $\Zt^{ij}$ as :
\[
  \Et^{ij} (\Zt^{kl}) = -\gO[i{[}k]\gO[l{]}j] -\ft18 \gO[ij]\gO[kl]
\]
and $\Ht$ by dilatations
\[
 \Ht (\Zt^{ij} ) = \Zt^{ij} \,.
\]
The  " special conformal generators" $\Ft^{ij}$ in the $\rep{\overline{27}}$ are realized nonlinearly:
\[
\Ft^{ij} (\Zt^{kl}) &:=& -2\,\Zt^{ij}(\Zt^{kl})
    + \gO[i{[}k]\gO[l{]}j] (\Zt^{mn}\Zt_{mn})
    +\ft18\, \gO[ij]\gO[kl] (\Zt^{mn}\Zt_{mn})\non[1ex]
&&{}+8\,\Zt\oversym{^{km}\Zt_{mn}\gO[n{[}i] \gO[j{]}l]}
    - \gO[kl](\Zt^{im}\gOd[mn]\Zt^{nj})
\lb{e7-nonlin-2}
\]
The norm form needed to define the $\EE$ invariant ``light cones'' is  constructed from the cubic invariant of
$\EEE$.  If we define the "distance" between $\Xt$ and $\Yt$ as  $\cN_3 (\Xt-\Yt)$ then it is manifestly
invariant under $\EEE$ and under the translations $\Et^{ij}$. Under $\Ht$ it transforms by a constant factor,
whereas under the action of $\Ft^{ij}$ we have
\[
 \Ft^{ij} \Big(\cN_3(\Xt - \Yt) \Big) =
  (\Xt^{ij} + {\Yt}^{ij})  \cN(\Xt - \Yt)\,.
\lb{e7-nonlin-3}
\]
which proves that the light cone in $\Rn^{27}$ with base point $\Yt$ defined by
\[
\cN_3(\Xt - \Yt) = 0
\]
is indeed invariant under $\EE$.

The above formulas carry over in a straightforward manner to the  conformal realization of $E_{7(-25)}$ on a 27
dimensional space coordinatized by the real exceptional Jordan algebra $J_3^{\mathbb{O}}$. In this case the cubic
form is invariant under $E_{6(-26)}$ which has $USp(6,2)$ as a subgroup. The $USp(8)$ covariant formulas above
for $\EE$ are then replaced by $USp(6,2)$ covariant formulas \cite{gupa}.

The conformal groups $\EE$ and $E_{7(-25)}$ acting on the 27 dimensional charge spaces of the $N=8$ and the
exceptional $N=2$ supergravity in five dimensions are isomorphic to the U-duality groups of the corresponding
four dimensional theories obtained by dimensional reduction. One may wander whether there exist conformal groups
acting on the charge space of  four dimensional supergravity theories that are isomorphic to the U-duality groups
of the corresponding three dimensional theories obtained by dimensional reduction. This question was investigated
in \cite{gkn1} and it was found that in the case of maximal supergravity,  even though there is no conformal
action of $E_{8(8)}$, it has a quasi-conformal group action on a 57 dimensional space which is an extension of
the 56 dimensional charge space by an extra coordinate. For BPS black hole solutions in $d=4$ this extra
coordinate can be taken to be the entropy \cite{gkn1}.

The realization of quasi-conformal action of $\E$ uses the 5-graded decomposition of its Lie algebra  with
respect to the Lie algebra of its $E_{7(7)}\times \cD$ subgroup
\[
\ba{c@{\,\,\oplus\,\,}c@{\,\,\oplus\,\,}c@{\,\,\oplus\,\,}c@{\,\,\oplus\,\,}c}
\fg^{-2} & \fg^{-1} & \fg^0 & \fg^{+1} &\fg^{+2}  \\[1ex]
\rep{1}  & \rep{56} & (\rep{133}\oplus\rep{1}) & \rep{56} & \rep{1} \ea \lb{e8-grading}
\]
with $\cD$ representing dilatations, whose generator together with grade $\pm 2$ elements generate an $SL(2,\Rn)$
subgroup. It turns out to be very convenient to work in a basis covariant with respect to the \SLR subgroup of
$\EE$ \cite{gkn1}. Let us denote the \SLR covariant generators belonging to the grade $-2,-1,0,1$ and $2$
subspaces in the above decomposition as follows:
\[
E \oplus \{E^{ij},E_{ij}\} \oplus \{G^{ijkl},\, G^i{}_j \, ;\, H\}
  \oplus \{F^{ij},F_{ij}\} \oplus F
\lb{e8-grading-gen}
\]
where $i,j,..=1,2...,8$ are now \SLR indices.

Consider now a  $57$-dimensional real vector space with coordinates
\[\cX:=(X^{ij},X_{ij},x) \] where $X^{ij}$ and $X_{ij}$ transform in the $28$ and $\tilde{28}$ of \SLR
and  $x$ is a singlet.   The generators of $\EE$ subalgebra act linearly on this space
\[
\begin{array}{rclrcl}
G^i{}_j(X^{kl}) &=& 2\,\gd\oversym{{}^k_j X^{il}} -\ft{1}{4}\gd^i_j X^{kl}\,, & G^{ijkl}(X^{mn})&=&
\ft{1}{24}\eps[ijklmnpq] X_{pq} \,,
\\[1ex]
G^i{}_j(X_{kl}) &=& -2\,\gd\undersym{{}^i_k X_{jl}}+\ft{1}{4}\gd^i_j X_{kl}\,, & G^{ijkl}(X_{mn})&=&
\gd^{[ij}_{mn} X^{kl]}_{\phantom{m}} \,,
\\[1ex]
G^i{}_j(x)    &=& 0 \,, & G^{ijkl}(x)   &=& 0 \,,
\end{array}\lb{e8-G}
\]
The generator $H$ of  dilatations  acts as
\[
\begin{array}{rclcrclcrcl}
H (X^{ij}) &=&  X^{ij} \,, &\quad& H (X_{ij}) &=&  X_{ij} \,, &\quad& H (x)    &=& 2\, x \,,
\end{array}\lb{e8-H}
\]
and the generator $E$  acts as translations on $x$:
\[
\begin{array}{rclcrclcrcl}
E(X^{ij}) &=& 0 \,, &\quad& E(X_{ij}) &=& 0 \,, &\quad& E(x) &=& 1 \,.
\end{array}\lb{e8-E}
\]
The grade $\pm 1$ generators act as
\[
\begin{array}{rclcrclcrcl}
E^{ij}(X^{kl}) &=& 0 \,, &\quad& E^{ij}(X_{kl}) &=& \gd{}^{ij}_{kl}  \,, &\quad&
E^{ij}(x)    &=& - X^{ij} \,, \\[1ex]
E_{ij}(X^{kl}) &=& \gd{}^{kl}_{ij} \,, &\quad& E_{ij}(X_{kl}) &=& 0 \,, &\quad& E_{ij}(x)    &=& X_{ij}\,.
\end{array}\lb{e8-Eij}
\]
The positive grade generators are realized nonlinearly. The generator $F$ acts as
\[
F(X^{ij}) &=&
 4 X\oversym{{}^{ik} X_{kl} X^{lj}}
 +X^{ij} X^{kl} X_{kl} \non
&&
 -\ft{1}{12}\eps[ijklmnpq] X_{kl} X_{mn} X_{pq}
 +X^{ij}\,x \,, \non[1ex]
F(X_{ij})  &=&
 -4 X\undersym{{}_{ik} X^{kl} X_{lj}}
 -X_{ij} X^{kl} X_{kl} \non
&&
 +\ft{1}{12}\epsd[ijklmnpq] X^{kl} X^{mn} X^{pq}
 +X_{ij}\,x \,, \non[1ex]
F(x) &=& 4\,\cI_4( X^{ij}, X_{ij})+x^2
\]
where $\cI_4$ is the quartic invariant of $\EE$
\[  \cI_4& \=
   X^{ij} X_{jk} X^{kl} X_{li}
 -\frac{1}{4} X^{ij} X_{ij} X^{kl} X_{kl}
+\ft{1}{96}\,\eps[ijklmnpq] X_{ij} X_{kl} X_{mn} X_{pq} \non
  &+\ft{1}{96}\,\epsd[ijklmnpq] X^{ij} X^{kl} X^{mn} X^{pq}
  \,\lb{e8-F}
\]
 The action of the remaining generators of $\E$ are as follows:
\[
F^{ij}(X^{kl}) &=&
 -4\, X\oversym{{}^{i[k} X^{l]j}}
 +\ft{1}{4}\,\eps[ijklmnpq] X_{mn} X_{pq} \,, \non[1ex]
F^{ij}(X_{kl}) &=&
 +8\,\gd\undersym{{}^{[i}_k X^{j]m}_{\phantom{m]n}} X_{ml}}
 +\gd{}^{ij}_{kl}\, X^{mn} X_{mn}   +2\, X^{ij} X_{kl}
 -\gd{}^{ij}_{kl}\,x \,, \non[1ex]
F_{ij}(X^{kl}) &=&
 -8\,\gd\oversym{{}_{[i}^k X_{j]m}^{\phantom{m}} X^{ml}}
 +\gd{}^{kl}_{ij}\, X^{mn} X_{mn}   -2\, X_{ij} X^{kl}
 -\gd{}^{kl}_{ij}\,x \,, \non[1ex]
F_{ij}(X_{kl}) &=&
 4\, X\undersym{{}_{ki} X_{jl}}
 -\ft{1}{4}\,\epsd[ijklmnpq] X^{mn} X^{pq} \,, \non[1ex]
F^{ij}(x) &=&
 4\, X\oversym{{}^{ik} X_{kl} X^{lj}} + X^{ij} X^{kl} X_{kl} \non
&&-\ft{1}{12}\,\eps[ijklmnpq]  X_{kl} X_{mn} X_{pq}
 + X^{ij}\,x \,, \non[1ex]
F_{ij}(x) &=&
 4\, X\undersym{{}_{ik} X^{kl} X_{lj}} + X_{ij} X^{kl} X_{kl} \non
&&-\ft{1}{12}\,\epsd[ijklmnpq]  X^{kl} X^{mn} X^{pq}
 - X_{ij}\,x  \,. \lb{e8-Fij}
\]

The above action of $\E$ was called quasiconformal in \cite{gkn1} since it leaves a certain norm invariant up to
an overall factor. Since the standard difference $ (\cX - \cY) $ of two vectors in the 57 dimensional space is
not invariant under "translations" generated by $(E^{ij},E_{ij})$, one defines a nonlinear difference that is
invariant under these  translation  as \cite{gkn1}
\[
\gd(\cX,\,\cY) \;:=\; (X^{ij}-Y^{ij},X_{ij}-Y_{ij}\;;\;x-y+\Sympl{X}{Y})\, = -\delta(\cY,\cX) \lb{difference}
\]
where $\Sympl{X}{Y}:= X^{ij}Y_{ij} -X_{ij}Y^{ij} $. One  defines  the norm of a vector $\cX$ in the 57
dimensional space as

\[
 \cN_4(\cX)\;\=\;\cN_4( X^{ij}, X_{ij}; x) \;:=\; \cI_4( X) - x^2\,, \lb{N4}
\]
Then the "distance" between any two vectors $\cX$ and $\cY$ defined as $ \cN_4(\gd(\cX,\cY)) $ is invariant under
$E_{7(7)}$ and translations generated by $E^{ij} , E_{ij}$ and $E$. Under the action of the remaining generators
of $\E$ one finds that
\[
F\Big(\cN_4(\gd(\cX,\cY))\Big) &=& 2\,(x+y)\,\cN_4(\gd(\cX,\cY)) \non F^{ij}\Big(\cN_4(\gd(\cX,\cY))\Big) &=&
  2\,(X^{ij}+Y^{ij})\,\cN_4(\gd(\cX,\cY)) \non
H\Big(\cN_4(\gd(\cX,\cY))\Big) &=& 4\,\cN_4(\gd(\cX,\cY)) \nn
\]
Therefore, for every $\cY\in \Rn^{57}$ the ``light cone'' with base point $\cY$, defined by the set of
$\cX\!\in\Rn^{57}$ satisfying
\[
\cN_4(\gd(\cX,\cY)) = 0\,,  \lb{E8lc}
\]
is preserved by the full $\E$ group.

The  quasiconformal realization of the other real noncompact form $E_{8(-24)}$ with the maximal compact subgroup
$E_7 \times SU(2)$ is given in  \cite{gupa}. In going to $E_{8(-24)}$ the subgroup $SL(8,\mathbb{R})$ of
$E_{7(7)}$ is replaced by the subgroup $SU^*(8)$ of $E_{7(-25)}$. The quasiconformal groups $E_{8(8)}$ and
$E_{8(-24)}$ are isomorphic to the U-duality groups of the maximal $N=16$ supergravity and the $N=4$ exceptional
supergravity in three dimensions. Since their action changes the "norm" in the charge-entropy space of the
corresponding four dimensional theories they can be interpreted as spectrum generating symmetry groups.

The quasiconformal realizations of $\E$ and $E_{8(-24)}$ can be consistently truncated to quasiconformal
realizations of other exceptional subgroups. For a complete list of the real forms of  these exceptional
subgroups we refer the reader to \cite{gkn1,gupa}.

 \section{ The minimal unitary representations of exceptional groups}

 To obtain  unitary realizations  of exceptional groups ,based on their quasiconformal realizations,
over certain  Hilbert spaces of square integrable functions one has to find the corresponding phase space
realizations of their generators and quantize them.  For $\E$ this was done in \cite{gkn2} and for $E_{8(-24)}$
in \cite{gupa}. Remarkably, the quantization of the quasiconformal realizations of $\E$ and $E_{8(-24)}$ yield
their minimal unitary representations. The concept of a minimal unitary representation of a non-compact group $G$
was first introduced by A. Joseph \cite{Joseph} and is defined as a unitary representation on a Hilbert space of
functions depending on the minimal number of coordinates for a given non-compact group.
 Here we shall summarize the results mainly for $\E$ and indicate how they
extend to $E_{8(-24)}$.

Since the positive graded generators form an Heisenberg algebra one introduces 28 coordinates $X^{ij}$ and 28
momenta $P_{ij}\equiv X_{ij}$, and one extra real coordinate $y$ to represent the central term. By quantizing
\[
[X^{ij},P_{kl}] \;=\; \i
\]
we can realize the positive grade generators of $\E$ as
\[
E^{ij} \;:=\; y\,X^{ij}\,, \quad E_{ij} \;:=\; y\,P_{ij}\,, \quad E\;:=\;\ft12\,y^2.
\]
To realize the other generators of $\E$ one introduces a momentum conjugate to the coordinate $y$ representing
the central charge of the Heisenberg algebra:
\[
 [y,p] \;=\; \i
\]
Then the remaining generators are given by
\[ \lb{e7e8gen}
 H &:=& \ft{1}{2} (y\,p+p\,y) \,,\non[2ex] F^{ij} &:=& -p\,X^{ij} + 2\i y^{-1}\, \Com{X^{ij}}{I_4(X,P)}\non
       &=&  -4\,y^{-1} X\oversym{^{ik}P_{kl}X^{lj}}
            - \ft{1}{2}\,y^{-1}(X^{ij}P_{kl}X^{kl}+X^{kl}P_{kl}X^{ij})\non
       &&
         +\ft{1}{12} y^{-1} \epsilon^{ijklmnpq} P_{kl}P_{mn}P_{pq}
         -p\,X^{ij}\,,\non[2ex]
F_{ij} &:=& -p\,P_{ij} + 2\i y^{-1}\, \Com{P_{ij}}{I_4(X,P)}\non
       &=&  4\,y^{-1} P\undersym{_{ik}X^{kl}P_{lj}}
            +\ft12\,y^{-1}(P_{ij} X^{kl}P_{kl}+P_{kl}X^{kl} P_{ij})\non
       &&
          -\ft{1}{12} y^{-1} \epsilon_{ijklmnpq} X^{kl}X^{mn}X^{pq}
          -p\,P_{ij}\,,\non[2ex]
F      &:=& \ft12 p^2 + 2 y^{-2} I_4(X,P)    \non[2ex]
 G^i{}_j  &:=& 2\, X^{ik}P_{kj}
              +\ft{1}{4} X^{kl}P_{kl} \,\gd^i_j \,,\non[2ex]
G^{ijkl} &:=& -\ft{1}{2} X^{[ij} X^{kl]}
              +\ft{1}{48} \epsilon^{ijklmnpq} P_{mn} P_{pq} \,.
\]
The hermiticity of all generators is manifest. Here $I_4(X,P)$ is the fourth order differential operator
\[ \label{invgen}
I_4(X,P) &:=& -\ft12(X^{ij}P_{jk}X^{kl}P_{li}+P_{ij}X^{jk}P_{kl}X^{li})\non
         &&   +\ft18(X^{ij}P_{ij}X^{kl}P_{kl}+P_{ij}X^{ij}P_{kl}X^{kl})\non
         &&   -\ft{1}{96}\,\epsilon^{ijklmnpq} P_{ij}P_{kl}P_{mn}P_{pq} \non
         &&   -\ft{1}{96}\,\epsilon_{ijklmnpq} X^{ij}X^{kl}X^{mn}X^{pq}
               + \ft{547}{16} \,.
\]
and  represents the quartic invariant of $\EE$ because
\[ \label{quartic}
\Com{G^i{}_j}{I_4(X,P)} = \Com{G^{ijkl}}{I_4(X,P)} = 0 \,.
\]

The above unitary realization of $E_{8(8)}$ in terms of position and momentum operators ( Schr\"{o}edinger
picture) can be reformulated in terms of annihilation and  creation operators (oscillator realization) (
Bargman-Fock picture) \cite{gkn2}. The transition from the Schrodinger picture to the Bargmann-Fock picture
corresponds to going from the $SL(8,\mathbb{R})$ basis to the $SU(8)$ basis of $E_{7(7)}$ .

 The quadratic Casimir operator of $\E$ reduces to a number for the above realization and one can show that
 all the higher Casimir operators must also reduce to numbers as required  for an
irreducible unitary representation. Thus by exponentiating the above generators we obtain the minimal unitary
irreducible representation of $\E$ over the Hilbert space of square integrable complex functions in 29 variables.

In the minimal unitary realization of the other noncompact real form $E_{8(-24)}$ with the maximal compact
subgroup $E_{7} \times SU(2)$ given explicitly in \cite{gupa} the relevant 5-graded decomposition of its Lie
algebra $\mathfrak{e}_{8(-24)}$ is with respect to its subalgebra $\mathfrak{e}_{7(-25)} \oplus
\mathfrak{so}(1,1)$
\begin{equation}
 \mathfrak{e}_{8(-24)}  =
\begin{array}{ccccccccc}
   \mathbf{1} & \oplus & \mathbf{56} & \oplus & \left(\mathfrak{e}_{7(-25)} \oplus
\mathfrak{so}(1,1)  \right) & \oplus &
   \mathbf{56} & \oplus & \mathbf{1}
\end{array}
\end{equation}

The Schr\"{o}dinger picture for the minimal unitary representation of $E_{8(-24)}$ corresponds to working in the
$SU^*(8) $ basis of the $E_{7(-25)}$ subgroup.  The position and momentum operators  transform in the ${\bf 28}$
and ${\bf \tilde{28}}$ of this $SU^*(8)$ subgroup  and the above formulas for $\E$ carry over  to those of
$E_{8(-24)}$ with some subtle differences\cite{gupa}. The Bargmann-Fock picture for the minimal unitary
realization of $E_{8(-24)}$ in terms of annihilation and creation operators is obtained by going from the
$SU^*(8)$ basis to the $SU(6,2)$ basis of the $E_{7(-25)}$ subgroup of $E_{8(-24)}$.

One can obtain the minimal unitary realizations of certain subgroups of $E_{8(8)}$ and $E_{8(-24)}$ by truncating
their minimal realizations. However, we should stress that since the minimal realizations of $E_{8(8)}$
\cite{gkn2}and $E_{8(-24)}$ \cite{gupa} are nonlinear consistent truncations exist for only certain subgroups.
The exceptional subgroups of $\E$ that can realized as quasiconformal groups are listed in \cite{gkn1} and the
possible consistent truncations of the minimal unitary realizations of $\E$ and $E_{8(-24)}$ are given in
\cite{gupa}. The relevant subalgebras of $\mathfrak{e}_{8(-24)}$ and $\mathfrak{e}_{8(8)}$ are those that are
realized as quasi-conformal algebras , i.e those that have a 5-grading
\[
  \mathfrak{g}= \mathfrak{g}^{-2} \oplus \mathfrak{g}^{-1} \oplus \mathfrak{g}^{0} \oplus \mathfrak{g}^{+1} \oplus
  \mathfrak{g}^{+2} \nonumber
\]
such that $\mathfrak{g}^{\pm 2}$ subspaces are  one-dimensional and $\mathfrak{g}^{0}= \mathfrak{h} \oplus \Delta
$ where $\Delta$ is the generator that determines the 5-grading.. Hence they all
 have  an  $\mathfrak{sl}(2,   \mathbb{R})$  subalgebra  generated  by
 elements of  $\mathfrak{g}^{\pm 2}$  and the generator  $\Delta$.  For the truncated subalgebra,
  the quartic invariant $\mathcal{I}_4$ will now  be that  of a  subalgebra $\mathfrak{h}$ of
 $\mathfrak{e}_{7(-25)}$ or of $\mathfrak{e}_{7(7)}$. Furthermore, this subalgebra must act on the
 grade $\pm 1$ spaces via symplectic representation. Below  we give the
 main chain of such subalgebras \cite{gupa}
\begin{equation}
 \mathfrak{h} = \mathfrak{e}_{7(-25)} \supset \mathfrak{so}^\ast(12) \supset \mathfrak{su}(3,3)
      \supset \mathfrak{sp}(6, \mathbb{R}) \supset \oplus_{1}^3 \mathfrak{sp}( 2, \mathbb{R} )
\supset \mathfrak{sp}( 2, \mathbb{R} ) \supset \mathfrak{u}(1)
\end{equation}
Corresponding quasi-conformal subalgebras read as follows
\begin{equation}
  \mathfrak{g} = \mathfrak{e}_{8(-24)} \supset \mathfrak{e}_{7(-5)} \supset \mathfrak{e}_{6(2)} \supset
      \mathfrak{f}_{4(4)}
   \supset \mathfrak{so}(4,4)
   \supset \mathfrak{g}_{2(2)} \supset \mathfrak{su}(2,1)
\end{equation}
The corresponding chains for the other real form $\mathfrak{e}_{8(8)}$ are
\begin{equation}
\mathfrak{h}= \mathfrak{e}_{7(7)} \supset \mathfrak{so(6,6)} \supset \mathfrak{sl(6,\mathbb{R}} \supset
\mathfrak{sp}(6, \mathbb{R}) \supset \oplus_{1}^3 \mathfrak{sp}( 2, \mathbb{R} ) \supset \mathfrak{sp}( 2,
\mathbb{R} ) \supset \mathfrak{u}(1)
\end{equation}
\begin{equation}
 \mathfrak{g}= \mathfrak{e}_{8(8)} \supset \mathfrak{e}_{7(7)} \supset \mathfrak{e}_{6(6)} \supset
     \mathfrak{f}_{4(4)}
 \supset \mathfrak{so}(4,4)
 \supset \mathfrak{g}_{2(2)} \supset \mathfrak{su}(2,1)
\end{equation}

The minimal unitary realizations of $\mathfrak{e}_{8(8)}$ and of $\mathfrak{e}_{8(-24)}$ can also be consistently
truncated to unitary realizations of certain subalgebras that act as regular conformal algebras with a 3-grading.
For $\mathfrak{e}_{8(8)}$ we have the following  chain of consistent truncations to conformal subalgebras
$\mathfrak{conf}$:
\begin{equation}
\mathfrak{conf} = \mathfrak{e}_{7(7)} \supset \mathfrak{so(6,6)} \supset \mathfrak{sl(6,\mathbb{R}} \supset
\mathfrak{sp}(6, \mathbb{R}) \supset \oplus_{1}^3 \mathfrak{sp}( 2, \mathbb{R} ) \supset \mathfrak{sp}( 2,
\mathbb{R} )
\end{equation}

 The corresponding chain of consistent truncations to conformal subalgebras for
$\mathfrak{e}_{8(-24)}$ is
\begin{equation}
\mathfrak{conf}=\mathfrak{e}_{7(-25)} \supset \mathfrak{so}^\ast(12) \supset \mathfrak{su}(3,3)
      \supset \mathfrak{sp}(6, \mathbb{R}) \supset \oplus_{1}^3 \mathfrak{sp}( 2, \mathbb{R} )
\supset \mathfrak{sp}( 2, \mathbb{R} )
\end{equation}

\section*{Acknowledgements} I would like to thank the organizers of the First Nordstr\"{o}m Symposium for their kind
 hospitality .



\begin{thebibliography}{0}


\bibitem{crjusc} E.~Cremmer, B.~Julia and J.~Scherk, ``Supergravity
  Theory in 11 Dimensions'', {\it Phys. \ Lett. }, {\bf B76}, 409 (1978).
\bibitem{witten95} E. Witten, `` String Theory Dynamics in various
  Dimensions'', {\it Nucl. Phys.} {\bf B443 } , 85 (1995).
\bibitem{julia} B~Julia, `` Group Disintegrations'', in: {\it
  Superspace and Supergravity}, eds. S.W.Hawking and M. Rocek,
  Cambridge University Press, 1981.
\bibitem{huto}C. Hull and P.K. Townsend, `` Unity of Superstring
  Dualities'', {\it Nucl.Phys.} {\bf B438} , 109 ( 1995).
\bibitem{GST1} M. G\"{u}naydin, G. Sierra and P.K. Townsend, Nucl. Phys.
\textbf{B242} (1984), 244 ; Phys. Lett. \textbf{133B} (1983) 72.
\bibitem{GST3} M. G\"{u}naydin, G. Sierra and P.K. Townsend, Class. Quantum
Grav. \textbf{3} (1986) 763.

\bibitem{gz2003}M. G\"{u}naydin and M. Zagermann, " Unified Maxwell-Einstein and Yang-Mills-Einstein supergravity theories  in
five dimensions,'' {\it JHEP} {\bf 0307}, 023 (2003)

\bibitem{d5bh} S. Ferrara and R. Kallosh, {\it Phys. Rev.} {\bf D54} (1996) 5344; hep-th/9603090. For further
references on the subject see \cite{fema,fegu}.
\bibitem{fema} See S. Ferrara and J. Maldacena, "Branes, central charges and U-duality invariant BPS conditions",
{\it Class.Quant.Grav.} {\bf 15} 749 ( 1998) and the references therein.

\bibitem{fegu}S.~Ferrara and M.~Gunaydin,
``Orbits of exceptional groups, duality and BPS states in string theory,'' {\it Int.\ J.\ Mod.\ Phys.} {\bf A13},
2075 (1998)
\bibitem{d4bh} R. Kallosh and B. Kol, {\it Phys. Rev.} {\bf D53} (1996) 1525; hep-th/9602014. For further
references on the subject see \cite{fema,fegu}.
\bibitem{gkn1}M.~Gunaydin, K.~Koepsell and H.~Nicolai,
``Conformal and quasiconformal realizations of exceptional Lie groups,'' {\it Commun.\ Math.\ Phys.}  {\bf 221},
57 (2001)
\bibitem{seegkn} For references on the subject see \cite{gkn1}.
\bibitem{gupa} M. Gunaydin and O. Pavlyk, to appear.
\bibitem{gkn2}M.~Gunaydin, K.~Koepsell and H.~Nicolai,
``The Minimal Unitary Representation of $E_{8(8)}$,''{\it  Adv.\ Theor.\ Math.\ Phys.}  {\bf 5}, 923 (2002)
\bibitem{Joseph}
A.~Joseph,  "Minimal realizations and spectrum generating algebras",  {\it Commun. Math. Phys.} \textbf{36}, 325
(1974); "The minimal orbit in a simple Lie algebra and its associated maximal
  ideal", {\it Ann. Sci. Ec. Norm. Super., IV. Ser.} \textbf{9}, 1 (1976). For further references on the minimal unitary
representations of noncompact exceptional groups see \cite{gkn2}.

\end{thebibliography}
\end{document}